# Defect processes in Be$_{12}$X Beryllides

**M. L. Jackson[a,b], P. A. Burr[c], R. W. Grimes[a]**

[a] Centre for Nuclear Engineering, Department of Materials, Imperial College London, SW7 2AZ, UK.
[b] Culham Centre for Fusion Energy, Culham Science Centre, Abingdon, Oxfordshire, OX14 3DB, UK.
[c] School of Electrical Engineering, University of New South Wales, Kingsford, NSW, 2052, Australia.

**Abstract**

The stability of intrinsic point defects in Be$_{12}$X intermetallics (where X = Ti, V, Mo or W) are predicted using density functional theory simulations and discussed with respect to fusion energy applications. Schottky disorder is found to be the lowest energy complete disorder process, closely matched by Be Frenkel disorder in the cases of Be$_{12}$V and Be$_{12}$Ti. Antisite and X Frenkel disorder are of significantly higher energy. Small clusters of point defects including Be divacancies, Be di-interstitials and accommodation of the X species on two Be sites were considered. Some di-interstitial, divacancy and X$_{2Be}$ combinations exhibit negative binding enthalpy (i.e. clustering is favourable), although this is orientationally dependent. None of the Be$_{12}$X intermetallics are predicted to exhibit significant non-stoichiometry, ruling out non-stoichiometry as a mechanism for accommodating Be depletion due to neutron transmutation.

**Introduction**

The use of beryllium (Be) in structural applications has been mostly limited to that of an alloying element, owing to the extreme toxicity of Be dust produced in machining. It is, nonetheless, employed due to its low atomic mass and unique neutronic properties; most notably as a window for X-rays and a neutron reflector and moderator in fission reactors[1].

Recently, Be has been used as a first wall material in experimental nuclear fusion reactors [2,3], since it causes relatively small radiative losses in the event it contaminates the plasma during a transient event. In future reactors, Be is proposed as a neutron multiplying material for tritium ($^3$T) breeding[4–6], since Be exhibits a low threshold for the (n,2n) reaction when bombarded with fast neutrons[a]. In these applications, Be will be subject to temperatures of 600°C[5,7] and a high flux of neutrons resulting in the generation of $^3$T and helium (He) through transmutation reactions. Several studies have indicated that this will cause an unacceptable degradation of its mechanical and thermal properties[8–10], along with retention of a high $^3$T inventory, producing a radiological hazard[11,12].

Be-rich intermetallics have been proposed as an alternative to elemental Be for nuclear applications, since they maintain similar neutronic properties to pure Be but perhaps offer a significant advantage in terms of $^3$T retention and radiation tolerance[13,14]. In particular, the Be$_{12}$X series, where X is a transition metal, have proved promising, with studies showing that Be$_{12}$Ti and Be$_{12}$V have adequate neutronic properties for use as a multiplier[15]. In comparison to Be, however, the irradiation response of Be intermetallics has not yet been adequately characterised, although several studies have investigated the response of Be$_{12}$Ti, showing it to compare favourably to pure Be in terms of embrittlement, swelling and tritium retention[14].

---

[a] note: here by Be we mean $^9$Be, while acknowledging $^{10}$Be is present as a trace impurity.

Further work must be carried out to identify the fundamental processes occurring during radiation damage in Be intermetallics. In particular, a greater understanding is required of how the point defects generated in damage cascades interact and lead to macroscopic changes in the microstructure. Recent work by Allouche *et al.*[16] on the isomorphic Be$_{12}$W structure used density functional theory to simulate the interactions of vacancies and hydrogen, finding that the intermetallic exhibits a significantly greater Be vacancy formation enthalpy in comparison to pure Be.

The study reported here contributes to our understanding of the fundamental processes occurring during radiation damage of Be intermetallics by predicting the formation and migration of intrinsic defects. It builds upon previous work focused on impurity behaviour in pure Be [17,18].

After a brief description of the computational methodology, we review the crystal structure of the Be$_{12}$X series, identifying all stable interstitial sites for intrinsic defects. The formation energies of all intrinsic point defects are evaluated, identifying the mechanisms relevant to stoichiometric, nonstoichiometric and radiation damage conditions.

Computational methodology

Density functional theory (DFT) simulations were carried out using the Perdew, Burke and Ernzerhof scheme of the generalised gradient approximation (GGA) for the exchange-correlation functional[19]. Ultra-soft pseudo potentials with a consistent cut-off of 480 eV (converged to 10$^{-3}$ eV atom$^{-1}$) were used throughout. All simulations were performed using the CASTEP code[20].

Defect calculations were performed in supercells constructed from 2 x 2 x 2 full Be$_{12}$X unit cells containing 208 atoms. A high density of k-points, with spacing of approximately 0.3 nm$^{-1}$ was used for the integration of the Brillion Zone, following the Monkhost-Packing scheme[21]. This corresponds to k-point grids of 2 x 2 x 4 for defect calculations.

As these materials are metallic, density mixing and Methfessel-Paxton[22] cold smearing of bands were employed with a width of 0.1 eV. Calculations were not spin-polarised, and during defect and elastic calculations no symmetry constraints were applied. All parameters, including the k-point spacing were converged to at least $10^{-3}$ eV.atom$^{-1}$.

For atomic relaxation in defective cells, the energy convergence criterion for self-consistent calculations was set to $10^{-7}$ eV and that for the forces on atoms to less than 0.01 eVÅ$^{-1}$. The cell was relaxed the stress component less than 0.05 GPa.

### Crystallography

It has been established that the crystal structure of this family of materials exhibits tetragonal symmetry and spacegroup I$_4$/mmm[23]. Several studies also report Be$_{12}$Ti as being hexagonal with spacegroup P$_6$/mmm[24,25]. In fact, Gilliam *et al.*[23] suggested that the large disordered hexagonal cell (with dimensions a = 29.44 and c = 7.33 Å) originally proposed by Raeuchle[25] is a derivative of the smaller tetragonal structure. This was confirmed in our recent work[26].

Within the I$_4$/mmm structure, the transition metal occupies the *2a* lattice site (0,0,0) with 12 fold coordination by Be (see fig. 1). Be occupies three symmetrically distinct sites within the structure, here named *Be1*, *Be2* and *Be3*. The *Be1* site (¼,¼,¼) has *8f* symmetry, coordinated by 10 Be sites and two transition metals. The *Be2* site (x,0,0) has *8i* symmetry and is coordinated by 13 Be sites and one transition metal. Finally, the *Be3* site (x,½,0) has *8j* symmetry and is coordinated with 10 Be sites and two transition metal sites. Thus, the *Be2* and *Be3* sites exhibit special positions, so that different compounds with the same crystal structures have different position values. The experimental position values of these sites presented in table 1, together with the lattice parameters, provide a useful part validation of the simulated data. The simulated lattice volumes are ~ 2% below their experimental volumes, typical for DFT predictions on metallic and intermetallic systems when using the PBE exchange correlation terms[19].

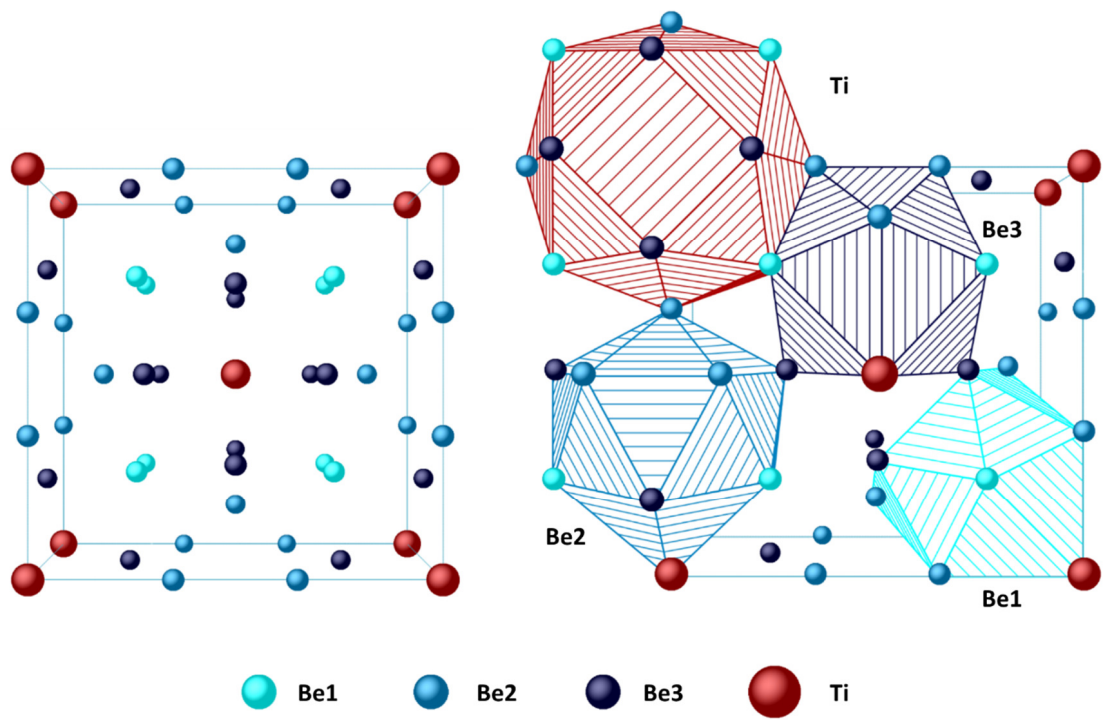

**Figure 1-** Left: position of stable Be and X interstitial sites within a unit cell of the I$_4$/mmm crystal structure of Be$_{12}$X, as viewed in the *c* direction. Right: coordination environment of symmetrically distinct lattice positions within the I$_4$/mmm crystal structure.

**Table 1 -** lattice parameters, volumes and special positions calculated for the Be$_{12}$X system, where X = V, Ti, Mo and W. x (Be2) and x (Be3) denotes the fractional x coordinate of these sites, which are special positions.

| Lattice properties | Be$_{12}$V Exp. | Calc. | Be$_{12}$Ti Exp. | Calc. | Be$_{12}$Mo Exp. | Calc. | Be$_{12}$W Exp. | Calc. |
|---|---|---|---|---|---|---|---|---|
| a (Å) | 7.266[a]-7.278[b] | 7.240 | 7.35[c] | 7.361 | 7.251[d]-7.271[b] | 7.239 | 7.340[e]-7.362[b] | 7.227 |
| c (Å) | 4.194[a]-4.212[b] | 4.169 | 4.19[c] | 4.163 | 4.232[d]-4.234[b] | 4.221 | 4.216[e]-4.232[b] | 4.223 |
| V (Å$^3$) | 221.4[a]-223.1[b] | 218.5 | 226.4[c] | 225.6 | 222.5[d]-223.8[b] | 221.2 | 227.1[e]-229.4[b] | 220.6 |
| x (Be2) | 0.361[a] | 0.349 | - | 0.350 | 0.351[b] | 0.350 | - | 0.351 |
| x (Be3) | 0.277[a] | 0.288 | - | 0.281 | 0.290[b] | 0.289 | - | 0.290 |

[a] reference [27], [b] reference [28], [c] reference [29], [d] reference [30], [e] reference [31]

While the perfect structure of this family of materials has been well characterised, interstitial sites within the I$_4$/mmm structure are identified here for the first time. This is achieved using the brute force approach described by Murphy[32], by seeding a Be$_{12}$Ti unit cell with a dense grid of Be and Ti interstitials with 0.03 nm spacing and performing a single point calculation to find the energy of these (unrelaxed) sites. The 20 lowest energy and symmetrically distinct defects were reproduced in a 2 x 2 x 2 supercells and geometry optimised to find the final position of the interstitial site. Four interstitial sites were identified as present within the I$_4$/mmm structure in fig. 2. Three interstitial sites can accommodate both Be and Ti, a *2b* site labelled *i1*, a *4b* site labelled *i2*, and an *8h* site labelled *i3*. A further site was found to be stable for the X species at (0,½,½) (*4c* symmetry) and is labelled *i4*. Be interstitials placed on this site move to a neighbouring *Be3* site, displacing the Be atom to the *i1* site when geometry optimised. In all cases, these present as typical interstitials in a complex structure, remaining on high symmetry sites and perturbing the surrounding lattice in a roughly symmetrical manner rather than forming a dumbbell as is common in simple metallic structures[33,34].

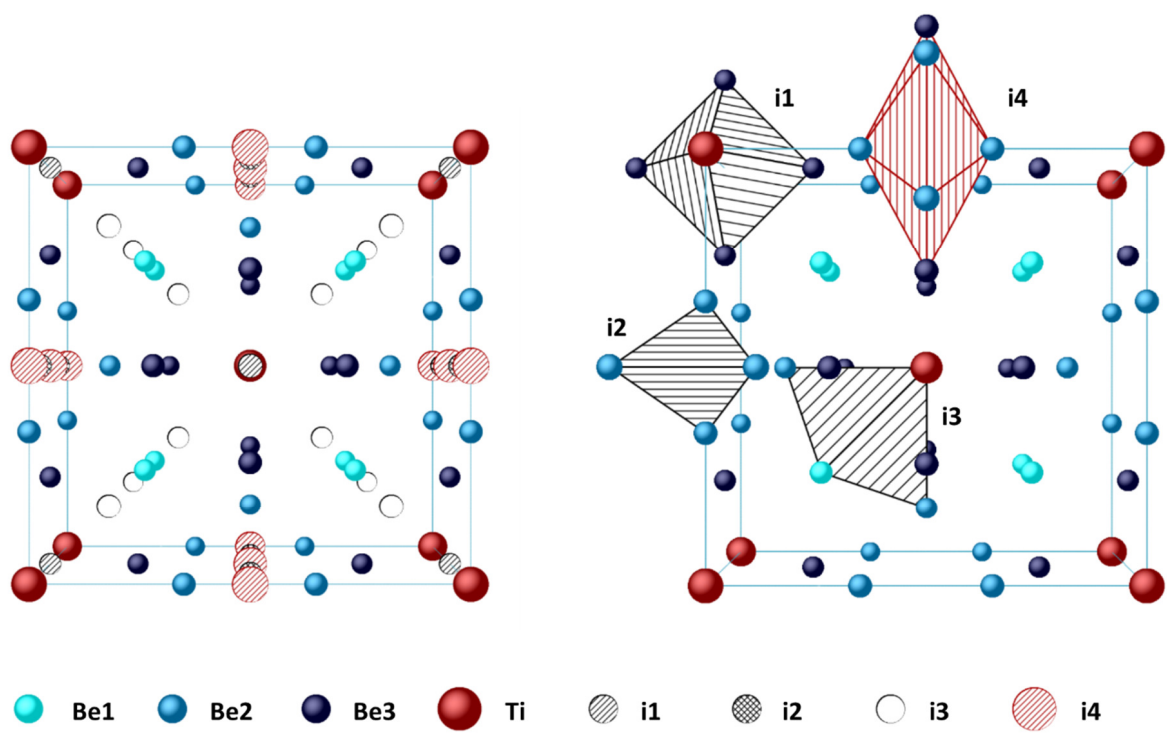

**Figure 2** - left: interstitial positions within a unit cell of the Be$_{12}$X structure as viewed along the *c* direction. Right: coordination environment of the four interstitial sites.

**Point Defects**

The formation enthalpies (E$_f$) of a Be vacancy and interstitial, denoted in Kröger-Vink notation[35] V$_{Be}$ and Be$_i$, are presented relative to their elemental reference states in table 2, following for example:

$$\text{Be}_{\text{Be}} \rightarrow \text{V}_{\text{Be}} + \text{Be}_{(s)}$$

**Table 2** - formation enthalpies, $E_f$, of Be vacancies and interstitials. $Be_{12}W$ is compared to previous DFT data. In bold, the most favourable defect of each type.

| | | | $E_f$/defect (eV) | | | | |
|---|---|---|---|---|---|---|---|
| Defect | $Be_{12}V$ | $Be_{12}Ti$ | $Be_{12}Mo$ | $Be_{12}W$ | $Be_{12}W^a$ | Defect | $Be^b$ |
| Be vacancies | | | | | | Be vacancies | |
| $V_{Be1}$ | 1.59 | 1.60 | 1.59 | 1.38 | 1.38 | $V_{Be}$ | 1.09 |
| **$V_{Be2}$** | **1.48** | **1.43** | **1.34** | **1.20** | **1.14** | Be interstitials | |
| $V_{Be3}$ | 1.64 | 1.53 | 1.66 | 1.47 | 1.48 | $Be_i$ (Oc) | 5.06 |
| Be interstitials | | | | | | $Be_i$ (Te) | 5.14 |
| $Be_{i1}$ | 2.95 | 3.19 | 3.54 | 3.81 | - | $Be_i$ (NBt) | 4.77 |
| **$Be_{i2}$** | **2.03** | **1.86** | **2.37** | **2.50** | - | $Be_i$ (Hx) | 5.67 |
| $Be_{i3}$ | 3.54 | 3.69 | 3.92 | 4.14 | - | $Be_i$ (Tr) | 4.01 |

a - Allouche *et al* (DFT data)[16], b - Middleburgh and Grimes (DFT data)[17]

For these materials, the $V_{Be2}$ site systematically exhibits the lowest $E_f$. $V_{Be1}$ has the next highest $E_f$, except for $Be_{12}Ti$ where the order of the *Be2* and *Be3* sites is reversed. For all cases there is only a small relative difference in $E_f$ between all three sites, reaching a maximum of 0.32 eV for $V_{Be2}$ and $V_{Be3}$ in $Be_{12}Mo$. For $Be_i$ the *i2* site systematically exhibits the lowest $E_f$, with the *i1* and *i3* sites significantly higher. This correlates to the proximity of the interstitial species to the X site: the *i2* site is surrounded only by Be whereas the *i1* and *i3* sites are also coordinated by an X atom. All $Be_i$ species have considerably higher $E_f$ than $V_{Be}$ species.

The formation energies of $V_X$ and $X_i$ are presented in table 3. Both $V_X$ and $X_i$ have significantly higher $E_f$ than $V_{Be}$ and $Be_i$. For all materials $V_X$ exhibit lower $E_f$ than the lowest enthalpy interstitial site, $X_{i4}$.

**Table 3** - formation enthalpies of transition metal vacancies and interstitials, presented in order of increasing transition metal radii. In bold, the lowest energy interstitial.

| | | $E_f$/defect (eV) | | | |
|---|---|---|---|---|---|
| Defect | $Be_{12}V$ | $Be_{12}Ti$ | $Be_{12}Mo$ | $Be_{12}W$ | $Be_{12}W^a$ |
| X vacancies | | | | | |
| $V_X$ | 3.37 | 4.10 | 3.61 | 3.16 | 3.25 |
| X interstitials | | | | | |
| $X_{i1}$ | 4.81 | 5.37 | 7.26 | 8.11 | - |
| $X_{i2}$ | 4.79 | 5.10 | 5.60 | 6.48 | - |
| $X_{i3}$ | 5.59 | 7.47 | 8.80 | 10.11 | - |
| **$X_{i4}$** | **4.69** | **4.19** | **4.84** | **5.95** | - |

a - Allouche *et al*[16] (DFT data)

Another important defect process resulting from radiation damage is antisite disorder, the $E_f$ of which are presented in table 4. In all materials the lowest $E_f$ defect is $X_{Be2}$; in the case of $Be_{12}W$ values are notably larger.

Table 4 - defect formation enthalpies for antisite defects in the $Be_{12}X$ series, presented in order of increasing atomic radii of the transition metal species. In bold, the lowest energy antisite.

| | $E_f$/defect (eV) | | | |
|---|---|---|---|---|
| defect | $Be_{12}V$ | $Be_{12}Ti$ | $Be_{12}Mo$ | $Be_{12}W$ |
| Anti-site | | | | |
| $Be_X$ | 2.83 | 3.55 | 3.09 | 2.76 |
| $X_{Be1}$ | 3.10 | 3.26 | 4.13 | 4.43 |
| **$X_{Be2}$** | **0.99** | **0.95** | **1.56** | **3.81** |
| $X_{Be3}$ | 1.79 | 2.50 | 3.40 | 3.81 |

*Defect clustering*

Since larger defects such as voids might form through coalescence of smaller clusters, how defects interact, particularly whether there is a driving force for association, is a key indicator of how the microstructure of a material will evolve during irradiation. As an initial step towards cluster formation, the binding enthalpies $E_B$ of nearest neighbour vacancies and interstitials were calculated with respect to the lowest enthalpy isolated defect of each type, that is, two $V_{Be2}$ or two $Be_{i2}$ (see table 5). Also, the cluster $X_{2Be}$ was considered, formed from $X_{Be}$ and $V_{Be}$ (which provides more volume for the larger X atom than a single vacancy). In all cases the propensity to form a cluster is indicated as a negative binding energy, for example,

$$V_{Be2} + V_{Be2} \rightarrow V_{Be2}V_{Be2} \; (Out\ of\ plane)$$

which is the lowest energy orientation cluster incorporating two lowest energy isolated vacancies (where out of plane indicates the two vacancies are orientated out of the basal plane). Alternately,

$$V_{Be2} + V_{Be2} + Be_{Be1} + Be_{Be3} \rightarrow V_{Be1}V_{Be3} + 2Be_{Be2}$$

will yield the binding energy to form the $V_{Be1}V_{Be3}$ cluster from two (lowest energy) isolated $V_{Be2}$ defects (which is positive and thus the cluster is not stable).

**Table 5 - Binding energies (where negative means bound and positive means unstable) of Be and X vacancies with respect to V$_{Be2}$ and V$_x$. In bold, the most favourable divacancy cluster of each type.**

|  | E$_B$ (eV) | | | |
|---|---|---|---|---|
| Divcacancy | Be$_{12}$V | Be$_{12}$Ti | Be$_{12}$Mo | Be$_{12}$W |
| V$_{Be3}$V$_{Be3}$ (in plane) | 0.37 | 0.41 | 0.74 | 0.92 |
| V$_{Be3}$V$_{Be3}$ (out of plane) | 0.37 | 0.44 | 0.35 | 0.74 |
| V$_{Be2}$V$_{Be3}$ (in plane) | 0.04 | 0.30 | 0.70 | 0.46 |
| V$_{Be2}$V$_{Be3}$ (out of plane) | -0.08 | 0.04 | 0.12 | 0.19 |
| V$_{Be2}$V$_{Be2}$ (in plane) | -0.01 | 0.22 | 0.03 | 0.11 |
| V$_{Be2}$V$_{Be2}$ (in plane) | -0.08 | 0.35 | 0.12 | 0.22 |
| **V$_{Be2}$V$_{Be2}$ (out of plane)** | **-0.21** | **-0.04** | **-0.09** | **-0.02** |
| V$_{Be2}$V$_{Be1}$ | -0.04 | 0.26 | 0.25 | 0.33 |
| V$_{Be1}$V$_{Be1}$ | 0.23 | 0.38 | 0.56 | 0.65 |
| V$_{Be1}$V$_{Be3}$ | 0.16 | 0.63 | 0.46 | 0.52 |
| V$_X$V$_X$ | 0.75 | 0.50 | 0.58 | 0.72 |
| V$_X$V$_{Be3}$ | -0.04 | -0.04 | 0.17 | 0.22 |
| **V$_X$V$_{Be2}$** | **-0.54** | **-0.41** | **-0.29** | **-0.19** |
| V$_X$V$_{Be1}$ | 0.04 | -0.02 | 0.37 | 0.42 |

Table 5 shows that for all materials the strongest cluster association by far is between V$_{Be2}$ and V$_X$, while two V$_X$ defects are strongly repelled in all materials.

Values of E$_B$ for two Be$_i$ on nearest neighbour sites are presented in table 6. Be$_i$Be$_i$ clusters include those on the *i4* site, despite this site not being stable for individual Be$_i$ defects. For most materials and interstitial combinations, E$_B$(Be$_i$Be$_i$) is repulsive, thus there is usually no driving force to form such clusters. One exception to this is E$_B$(Be$_{i4}$Be$_{i4}$) which is slightly negative in both Be$_{12}$Ti and Be$_{12}$V and only moderately repulsive in Be$_{12}$Mo and Be$_{12}$W. Thus, we predict no driving force for the formation of larger interstitial defect clusters, through a model where isolated interstitial defects associate initially into pairs, other than via two interstitials reorienting onto i4 sites.

**Table 6- Binding enthalpies (where negative means bound and positive means unstable) of Be$_i$ Be$_i$ with respect to two Be$_{i2}$. In Bold, the most favourable di-interstitial cluster.**

|  | Binding enthalpy (eV) | | | |
|---|---|---|---|---|
| Interstitial sites | Be$_{12}$V | Be$_{12}$Ti | Be$_{12}$Mo | Be$_{12}$W |
| Be$_{i2}$Be$_{i2}$ | 0.38 | 0.71 | 0.37 | 0.42 |
| Be$_{i3}$Be$_{i1}$ (in plane) | 1.90 | 2.41 | 2.18 | 2.32 |
| Be$_{i3}$Be$_{i1}$ (out plane) | 1.98 | 2.50 | 2.10 | 2.22 |
| Be$_{i3}$Be$_{i2}$ | 0.34 | 0.17 | 0.69 | 0.71 |
| Be$_{i3}$Be$_{i3}$ | 1.69 | 2.50 | 1.82 | 1.86 |
| Be$_{i3}$Be$_{i4}$ (in plane) | 1.74 | 2.17 | 2.01 | 2.16 |
| Be$_{i3}$Be$_{i4}$ (out plane) | 1.96 | 2.42 | 0.69 | 0.71 |
| Be$_{i4}$Be$_{i2}$ | 0.43 | 0.38 | 0.32 | 0.38 |
| **Be$_{i4}$Be$_{i4}$** | **-0.10** | **-0.10** | **0.08** | **0.09** |

The binding energies for $V_{Be}$ and $X_{Be}$ to form $X_{2Be}$ are presented in table 7. $E_B$ ranges from strongly negative (-4.55 eV) to strongly positive (4.26 eV). The lowest energy sites are split between *Be2-Be3* for $Be_{12}V$ and $Be_{12}Mo$, and *Be2-Be2* for $Be_{12}Ti$ and $Be_{12}W$. The latter case might have been anticipated, as this position corresponds roughly to the arrangement of atoms in the $Be_{17}Ti_2$ Pc/mmm phase, which is closely related to the $Be_{12}Ti$ $I_4$/mmm phase[23].

Table 7 - Binding enthalpy (where negative means bound and positive means unstable) of $X_{2Be}$ with respect to $X_{Be2}$ and $V_{Be2}$. In bold, the most favourable cluster for each material.

|  | Binding enthalpy (eV) | | | |
| --- | --- | --- | --- | --- |
| Anti-site vacancy pair | $Be_{12}V$ | $Be_{12}Ti$ | $Be_{12}Mo$ | $Be_{12}W$ |
| $X_{Be3-Be3}$ (in plane) | 0.33 | 1.35 | 0.63 | -1.77 |
| $X_{Be3-Be3}$ (out of plane) | -0.42 | 0.17 | -0.16 | -1.28 |
| $X_{Be2-Be3}$ (in plane) | -0.51 | -0.09 | -0.45 | -1.48 |
| **$X_{Be2-Be3}$ (out of plane)** | **-2.46** | -0.91 | **-2.88** | -2.04 |
| $X_{Be2-Be2}$ (in plane) | -0.40 | 4.26 | -0.02 | -1.77 |
| $X_{Be2-Be2}$ (in plane) | 3.44 | 0.00 | -0.32 | -1.67 |
| **$X_{Be2-Be2}$ (out of plane)** | -0.40 | **-3.02** | 0.17 | **-4.55** |
| $X_{Be2-Be1}$ | -0.06 | -0.38 | -0.16 | -2.22 |
| $X_{Be1-Be3}$ | 0.75 | -0.46 | 0.54 | -1.93 |
| $X_{Be1-Be2}$ | 0.75 | 0.00 | 0.03 | -1.01 |

**Defect Disorder Processes**

Defects can be generated in a material through several different processes (disorder reactions) that can occur thermally, or be driven by radiation damage cascades. Table 8 shows the energies associated with Frenkel, Schottky and Antisite processes (normalised by the number of defects for each process).

Table 8 - Normalised defect process formation energies, $E_f$, for Frenkel, Schottky and (simple) antisite disorder.

|  | $E_f$/defect (eV) | | | |
| --- | --- | --- | --- | --- |
| Defect | $Be_{12}V$ | $Be_{12}Ti$ | $Be_{12}Mo$ | $Be_{12}W$ |
| Be Frenkel | 1.76 - 2.59 | 1.64 – 2.65 | 1.85 – 2.79 | 1.85 - 2.81 |
| X Frenkel | 4.03 – 4.48 | 4.15 – 5.78 | 4.22 – 6.20 | 4.55 – 6.63 |
| Schottky | 1.63 – 1.77 | 1.63 – 1.79 | 1.51 – 1.81 | 1.35 – 1.60 |
| Antisite | 1.91 – 2.96 | 2.25 – 3.41 | 2.33 – 3.61 | 3.29 – 3.60 |

The range of $E_f$ values for vacancy, interstitial and anti-site defects leads to a range of values for each disorder process, which can vary by as much as 2 eV or as little as 0.2 eV. While the minimum value could be the most significant in equilibrium processes, radiation damage is not an equilibrium process and thus higher enthalpy configurations may also be important.

For all of these materials, Schottky disorder is the lowest enthalpy disorder process, while Be Frenkel disorder exhibits a similar albeit slightly higher energy. Indeed for $Be_{12}Ti$, Schottky and Frenkel disorder energies are essentially identical. Conversely, X Frenkel disorder is a much higher energy process in all materials. Thus, we predict a strong thermodynamic driving force for the removal of $X_i$ species from a damaged lattice. In the case of $Be_{12}V$, antisite disorder will also be significant, being within 0.3 eV/defect of the Schottky process.

**Nonstochiometry**

Due to the defect disorder processes examined previously, deviation from stoichiometry may occur. This has been assessed by calculating the energy to dissolve a formula unit of the nearest 0 K reference state into $Be_{12}X$, creating defects in the $Be_{12}X$ lattice. For instance in the case of $Be_{12}Ti$, where $Be_{17}Ti_2$ phase incorporation results in $V_{Be}$ formation:

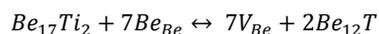

$$Be_{17}Ti_2 + 7Be_{Be} \leftrightarrow 7V_{Be} + 2Be_{12}T$$

where $Be_{Be}$ are Be atoms on Be sites in the host $Be_{12}Ti$ lattice. A full list of these equations and reference states is presented in the appendix. The minimum energy for the incorporation of reference states is shown in table 9. The number of defects formed can then be calculated at temperature using the Arrhenius approximation and total deviation from stoichiometry calculated, as presented in fig. 3.

Table 9 - **Solution energy to closest compositional reference state that results in the formation of a single defect and hence a change in stoichiomtry. Governing equations can be found in the appendix.**

|  | $E_f$/defect (eV) | | | |
| --- | --- | --- | --- | --- |
| Defect | $Be_{12}Ti$ | $Be_{12}V$ | $Be_{12}Mo$ | $Be_{12}W$ |
| $V_{Be}$ | 1.52 | 1.53 | 1.36 | 1.26 |
| $V_X$ | 3.34 | 3.23 | 1.46 | 1.41 |
| $Be_i$ | 2.49 | 2.34 | 2.23 | 2.37 |
| $X_i$ | 6.04 | 5.39 | 5.64 | 6.76 |
| $Be_X$ | 2.79 | 2.66 | 0.81 | 0.88 |
| $X_{Be}$ | 2.98 | 1.78 | 2.39 | 4.67 |
| $X_{2Be}$ | 1.48 | 0.84 | 0.88 | 1.38 |
| $V_{Be}V_{Be}$ | 2.82 | 2.97 | 2.64 | 2.49 |
| $V_XV_{Be}$ | 4.45 | 4.19 | 2.54 | 2.47 |
| $V_XV_X$ | 7.18 | 7.17 | 3.50 | 3.54 |
| $Be_iBe_i$ | 4.89 | 4.59 | 4.54 | 4.82 |

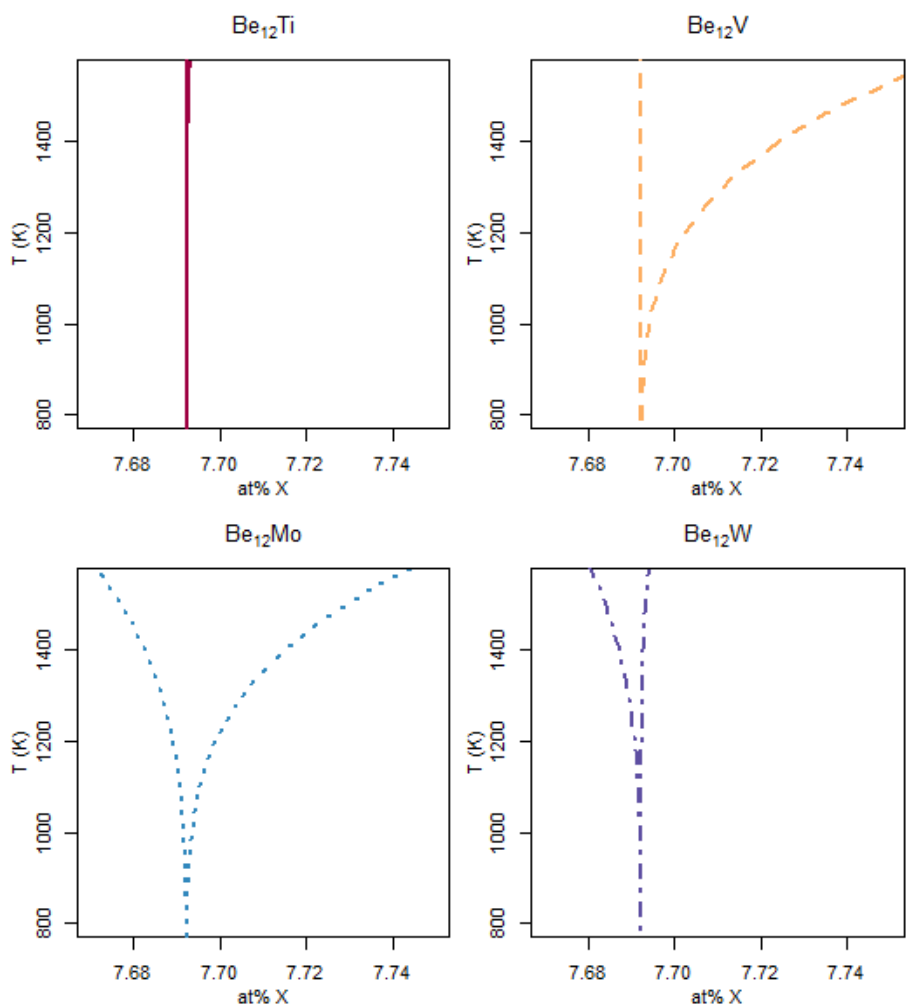

**Figure 3** – Predicted phase field lines based on total defect concentrations predicted through the Arrhenius approximation for materials with an excess of Be and the X species.

All compounds exhibit almost no deviation from stoichiometry (note that the full range of the ordinate axis corresponds to a compositional variation of 0.08 at%). This is especially true for $Be_{12}Ti$, while $Be_{12}V$ may accommodate limited Be sub-stoichiometry, $Be_{12}W$ may accommodate very minor levels of Be hyper-stochiometric, and $Be_{12}Mo$ may accommodate small deviation on both side of the stoichiometric composition. This difference in behaviour can be attributed to the relative energy of the three lowest energy defect reactions resulting in $V_{Be}$ $X_{Be}$ and $X_{2Be}$ formation. In $Be_{12}Ti$ the lowest energy defect is $Ti_{2Be}$ with energy 1.48 eV, considerably higher than for other materials thereby promoting little deviation from stoichiometry. In the case of $Be_{12}V$, $V_{2Be}$ has a lowest energy, 0.84 eV, so that any non-stoichiometry will be accommodated by an excess of the transition metal. A similar profile occurs in $Be_{12}Mo$, where $Mo_{2Be}$ has energy 0.88 eV, however $Be_{Mo}$ has formation energy 0.81 eV which also allows for some non-stochiometry in the Be rich region. $Be_{12}W$ becomes hyperstochiometric as $Be_W$ has energy 0.88 eV.

The extent of non-stochiometry predicted in these materials is such as they may be considered line compounds. Thus, upon depletion of Be due to the (n,2n) reaction, these materials are likely to form secondary phases with composition $Be_{17}V_2$, $Be_{17}Ti_2$, $Be_2W$ and $Be_2Mo$, to accommodate excess X. From this point of view, $Be_{12}V$ and $Be_{12}Ti$ are likely to be the least affect by the formation of secondary phases, not because of their ability to accommodate deveiations from stoicbiometry, but because the secondary phases have structural similarities to the parent phase[23] and not too dissimilar composition. Following the same logic, $Be_{22}W$ and $Be_{22}Mo$ should be considered as possible candidates over $Be_{12}W$ and $Be_{12}Mo$.

**Conclusions**

Density functional theory simulations have been carried out in order to predict defect properties of $Be_{12}X$ materials. Calculated perfect lattice parameters were in good agreement with experiment. Formation enthalpies of all possible vacancies, interstitials and anti-site point defects were calculated for all four materials. Four stable sites for intrinsic interstitials are identified for the first time in the $I_4/mmm$ structure: the *i2* site (*4b*) is most stable for Be self-interstitials but for $X_i$ species, the i4 (*4c*) site offers a similar or lower energy. In all cases, the Be sub-lattice accommodates defects more readily than the X sub-lattice. Formation energies of point defects were combined to predict the energies of intrinsic disorder processes. Of these, Schottky disorder was identified as the lowest energy process, while Be Frenkel disorder exhibited only slightly higher energy for the V and Ti containing materials.

Small clusters including $V_{Be}V_{Be}$, $Be_iBe_i$ and $X_{2Be}$ were investigated, with some combinations exhibiting favourable binding enthalpy, particularly for $X_{2Be}$, which in some cases has a notably lower enthalpy than that of simple antisite disorder. This is likely due to the large size discrepancy between Be and the X species.

$Be_{12}Mo$, $Be_{12}V$ and $Be_{12}W$ can only exhibit modest non-stoichiometry at elevated temperatures, while $Be_{12}Ti$ is essentially a true line compound due the relatively high energy required to form a defect relative to its nearest reference state. The ability to accommodate the removal of Be atoms (X excess) without severely affecting the compounds stability is a noteworthy property for neutron multiplier purposes, where Be atoms are continually consumed to maintain the necessary neutron flux. Consideration of which secondary phases might form is therefore important.

**Acknowledgments**

MLJ has received funding via CCFE from the Euratom research and training programme 2014-2018 under grant agreement No 633053 (EUROfusion Consortium) - the views and opinions expressed herein do not necessarily reflect those of the European Commission. Computing resources were provided by the Imperial College London High Performance Computing Service. Dr. Dmitry Bachurin is acknowledged for his useful insight on the possible importance of X species accommodation on a Be divacancy.

Appendix

**Table 12 - Reference states and defect equations evaluated to calculate non-stochiometry.**

| Material  Reference states | Be$_{12}$W and Be$_{12}$Mo  Be$_{22}$W/Be$_2$W, Be$_{22}$Mo/Be$_2$Mo | Be$_{12}$Ti and Be$_{12}$V  Be/Be$_{17}$Ti$_2$, Be/ Be$_{17}$V$_2$ |
|---|---|---|
| $V_{Be}$ | $Be_2X + 10Be_{Be} \leftrightarrow 10V_{Be} + Be_{12}X$ | $Be_{17}X_2 + 7Be_{Be} \leftrightarrow 7V_{Be} + 2Be_{12}X$ |
| $V_X$ | $6Be_{22}X + 5X_X \leftrightarrow 5V_X + 11Be_{12}X$ | $12Be + X_X \leftrightarrow V_X + Be_{12}X$ |
| $Be_i$ | $Be_{22}X \leftrightarrow 10Be_i + Be_{12}X$ | $Be_{17}X_2 + 8Be \leftrightarrow 2Be_{12}X + Be_i$ |
| $X_i$ | $6Be_2X \leftrightarrow 5X_i + Be_{12}X$ | $12Be_{17}X_2 \leftrightarrow 7X_i + 17Be_{12}X$ |
| $2V_{Be}$ | $Be_2X + 10Be_{Be} \leftrightarrow 5(2V_{Be}) + Be_{12}X$ | $2Be_{17}X_2 + 14Be_{Be} \leftrightarrow 7(2V_{Be}) + 4Be_{12}X$ |
| $2V_X$ | $12Be_{22}X + 5X_X \leftrightarrow 5(2V_X) + 22Be_{12}X$ | $24Be + 2X_X \leftrightarrow (2V_X) + 2Be_{12}Ti$ |
| $V_{Be}V_X$ | $11Be_{22}X + 10X_X + 10Be_{Be} \leftrightarrow 10V_{Be}V_X + 21Be_{12}X$ | $11Be + Be_{Be} + Ti_{Ti} \leftrightarrow V_{Be}V_{Ti} + Be_{12}Ti$ |
| $2Be_i$ | $Be_{22}X \leftrightarrow 5(2Be_i) + Be_{12}X$ | $2Be_{17}Ti_2 + 16Be \leftrightarrow 4Be_{12}Ti + (2Be_i)$ |
| $Be_X$ | $13Be_{22}X + 10X_X \leftrightarrow 10Be_X + 23Be_{12}X$ | $13Be + Be_{Be} + Ti_{Ti} \leftrightarrow Be_{Ti} + Be_{12}Ti$ |
| $X_{Be}$ | $13Be_2X + 10Be_{Be} \leftrightarrow 10X_{Be} + 3Be_{12}X$ | $2Be_{17}Ti_2 + Be_{Be} + Be \leftrightarrow Ti_{Be} + 3Be_{12}Ti$ |
| $X_{2Be}$ | $14Be_2X + 20Be_{Be} \leftrightarrow 10X_{2Be} + 4Be_{12}X$ | $2Be_{17}Ti_2 + 2Be_{Be} \leftrightarrow Ti_{2Be} + 3Be_{12}Ti$ |